\documentclass[sigconf,nonacm,balance=false]{acmart}
\setcopyright{none}
\AtBeginDocument{%
  }

\usepackage{graphicx}
\usepackage{subcaption}
\usepackage{booktabs}   % \toprule \midrule \bottomrule
\usepackage{multirow}
\usepackage{tabularx}
\usepackage{array}
\usepackage{xcolor}                 % color mixes like black!85, gray!5
\usepackage{listings}               % lstlisting in the appendix
\usepackage[most]{tcolorbox}        % tcblisting + skins + breakable, etc.
\tcbuselibrary{listings}
\usepackage{adjustbox}

\begin{document}

\title[GeoCloak: Defending Against Geolocation Reasoning]{GeoCloak: Defending Against Geolocation Reasoning in Multimodal Language Models via Perceptual Image Editing}

\author{Haoyang Xu}
\affiliation{\institution{University of Hawai`i at M\=anoa}\city{Honolulu}\country{USA}}
\email{haoyangx@hawaii.edu}

\author{Christian Iha}
\affiliation{\institution{University of Hawai`i at M\=anoa}\city{Honolulu}\country{USA}}
\email{ihac@hawaii.edu}

\author{Colbren Fujimoto}
\affiliation{\institution{University of Hawai`i at M\=anoa}\city{Honolulu}\country{USA}}
\email{colbren@hawaii.edu}

\author{Yingfei Dong}
\affiliation{\institution{University of Hawai`i at M\=anoa}\city{Honolulu}\country{USA}}
\email{yingfei@hawaii.edu}

\author{Rui Ning}
\affiliation{\institution{Old Dominion University}\city{Norfolk}\country{USA}}
\email{rning@cs.odu.edu}

\author{Liuwan Zhu}
\affiliation{\institution{University of Hawai`i at M\=anoa}\city{Honolulu}\country{USA}}
\email{liuwan@hawaii.edu}

\renewcommand{\shortauthors}{Xu et al.}

\begin{abstract}
Multimodal large language models can infer sensitive geographic information from ordinary images by combining visual, textual, architectural, and environmental cues. This capability creates a privacy risk: users may reveal fine-grained location information even when explicit metadata such as GPS tags has been removed. We study this problem as geolocation inference in multimodal language models and present GeoCloak, a cue-aware perceptual editing framework for reducing location leakage from shared images. GeoCloak uses an accessible multimodal model as a proxy to identify geolocation-relevant cues, localizes high-impact regions through object grounding, and applies localized semantic edits designed to disrupt downstream geolocation predictions while preserving the overall scene content. Experiments on street-level image datasets show that GeoCloak substantially reduces administrative-level geolocation accuracy across multiple multimodal models, while maintaining automatic perceptual-quality scores closer to those of the original images than pixel-level perturbation baselines. These results suggest that cue-level semantic editing is a promising direction for reasoning-based visual privacy, while also highlighting open challenges in deployment cost, semantic plausibility, and adaptive evaluation.
\end{abstract}

\keywords{Location privacy, multimodal language models, image geolocation, perceptual image editing}

\maketitle
\hypersetup{pdfauthor={Haoyang Xu, Christian Iha, Colbren Fujimoto, Yingfei Dong, Rui Ning, Liuwan Zhu}}

\section{Introduction}
\label{sec:intro}

Recent multimodal large language models (MLLMs) have shown strong capabilities in open-world visual reasoning, including scene understanding, OCR-based inference, visual question answering, and image geolocation. In geolocation tasks, these models can combine visual cues such as architecture, vegetation, road layout, and storefront appearance with textual cues such as signs, addresses, and license-plate patterns. As a result, ordinary public images may reveal sensitive location information even when explicit metadata such as GPS tags has been removed.

\begin{figure}[t]
    \centering
    \includegraphics[width=\linewidth]{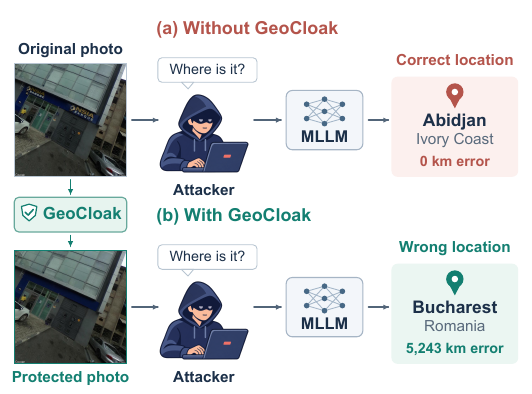}
    % 改成figure1
    % \caption{Example of GeoCloak defense. }
    \caption{Threat model and defense. (a) Without protection, an attacker queries an MLRM with a shared photo and recovers its location (Abidjan, 0\,km error). (b) After \textsc{GeoCloak} edits a localized geolocation cue, the same query yields a wrong location (Bucharest, 5,243\,km error) while the photo remains visually natural.}
    \label{fig:combined_result}\vspace{-0.5cm}
\end{figure}

This capability creates a practical privacy risk for user-shared images. With only a publicly shared photo, an API key, and access to mapping services, an adversary can launch an automated geolocation attack: prompting the model with “Where is this place?”, inspecting the model’s reasoning trace to extract high-information cues(e.g., license-plate style, street vegetation, architectural patterns), and cross-checking the predicted coordinates on Google Street View or OpenStreetMap. Recent work~\cite{luo2025doxing} shows that such attacks can achieve up to 99\% de-anonymization success rates, revealing users’ home addresses, daily routines, and sensitive locations from seemingly innocuous posts on social platforms.

Existing image privacy defenses are not designed for this form of reasoning-based leakage. Recent work on MLRM-based doxing~\cite{luo2025doxing} shows that even after explicit metadata is removed, multimodal models can still exploit residual visual and textual cues to infer fine-grained locations. Removing EXIF metadata eliminates explicit GPS tags, but does not affect the visual evidence used by multimodal models. Blurring or masking can hide known sensitive regions, but it requires knowing in advance which cues matter and may remove visually useful content. Pixel-level perturbations can reduce model accuracy, but they often degrade visual quality and may be fragile under resizing, compression, or denoising. In essence, most current methods focus on \emph{preventing the model from seeing}, whereas few address the harder challenge of \emph{allowing it to see but preventing it from reasoning correctly}. 

We propose GeoCloak, a cue-aware perceptual editing framework for reducing geolocation leakage from shared images. GeoCloak uses an accessible multimodal model as a proxy to identify geolocation-relevant cues, localizes high-impact regions, and applies localized semantic edits such as removing, replacing, or adding scene-consistent cues. As illustrated in Figure~\ref{fig:combined_result}, GeoCloak targets the visual evidence that supports geolocation inference, aiming to change model predictions while preserving most of the original scene content rather than globally corrupting the image.

Our key contributions are:
\begin{itemize}
    \item We formulate image geolocation by multimodal language models as a reasoning-based privacy risk. Building on recent evidence that MLLMs can infer fine-grained locations from visual and textual cues~\cite{luo2025doxing,huang2025vlms}, we analyze how such cues support location inference in shared images.

    \item We propose \textsc{GeoCloak}, a cue-aware perceptual editing framework for reducing geolocation leakage. \textsc{GeoCloak} uses an accessible multimodal model as a proxy to identify geolocation-relevant cues, localizes high-impact regions with object grounding~\cite{ren2024grounded}, and applies localized semantic edits to reduce downstream geolocation accuracy.

    \item We evaluate \textsc{GeoCloak} on two street-level geolocation datasets, Doxing~\cite{luo2025doxing} and FairLocator~\cite{huang2025vlms}, across multiple multimodal models. Results show that cue-level editing substantially reduces administrative-level geolocation accuracy while maintaining automatic perceptual-quality scores closer to those of the original images than pixel-level perturbation baselines.
\end{itemize}

\section{Related Work}

\begin{figure*}[t]
    \centering
    \includegraphics[width=0.8\linewidth]{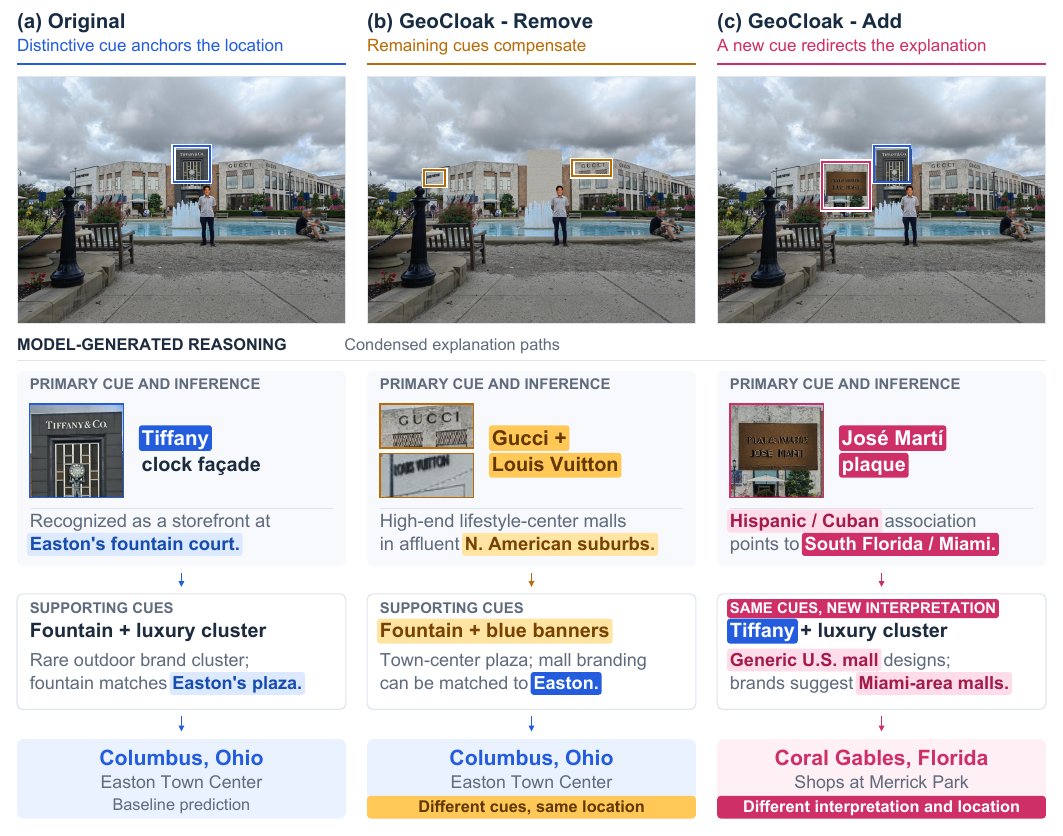}
    % \caption{An illustration of the experiment observation.}
    \caption{Empirical observation. (a) The model localizes the original image (Columbus, Ohio) from a primary cue plus supporting context. (b) Removing the primary cue does not change the prediction: the model falls back on other cues to keep a coherent explanation. (c) Adding a plausible new cue (a Jos\'e Mart\'i plaque) redirects both the explanation and the prediction (Coral Gables, Florida).}
    \label{fig:attention}
\end{figure*}
\subsection{Multimodal Large Reasoning Models.}
Multimodal Large Reasoning Models (MLRMs) mark the convergence of perception and reasoning in AI~\cite{achiam2023gpt, yin2024survey}. 
Evolving from Large Language Models (LLMs)~\cite{brown2020language} and Multimodal LLMs (MLLMs), these systems integrate visual perception with chain-of-thought reasoning~\cite{wei2022chain, yao2023react}. 
Recent MLRMs such as OpenAI’s \emph{o3} combine visual parsing, OCR, and world knowledge to perform fine-grained inference, including open-world geolocation~\cite{luo2025doxing}. 
While these capabilities unlock powerful applications, they also lower the barrier for extracting sensitive spatial information from ordinary photos, raising new challenges for location privacy protection.

\subsection{Geolocation Privacy Attacks and Defenses.}
The privacy implications of visual data have long been recognized~\cite{orekondy2017vispr}. Early works such as IM2GPS~\cite{hays2008im2gps} and PlaNet~\cite{weyand2016planet} achieved coarse retrieval-based localization, later refined by deep place-recognition systems~\cite{arandjelovic2016netvlad}. 
With CLIP-style models~\cite{geoclip} and multimodal large reasoning models (MLRMs)~\cite{luo2025doxing, liu2024lvln_geoloc}, localization accuracy has advanced toward street-level, enabling reasoning-driven privacy threats such as doxing. Beyond direct geolocation inference, Mendes et al.~\cite{mendes2024granular} investigate granular privacy control in VLM-based geolocation dialogues, highlighting the importance of regulating location disclosure at different geographic levels.

Concurrently, a growing body of work has explored adversarial image attacks to mislead vision–language and reasoning models: for example, evaluations of model robustness to adversarial images~\cite{dong2023robust}, diffusion-based methods that improve transferability to closed models~\cite{guo2024efficient}, large-scale self-supervised attack frameworks~\cite{zhang2025anyattack}, and surprisingly effective black-box baselines~\cite{li2025frustratingly}. Although these methods can yield privacy-preserving effects under certain conditions, many rely on low-resolution optimization or aggressive down-sampling, which substantially reduces image usability, and the resulting perturbed images often exhibit visible texture artifacts that make the manipulation perceptible to humans.
These shortcomings motivate defenses that intervene at the reasoning level (rather than merely obfuscating pixels) to more robustly protect location privacy. Diffusion-based approaches have also enabled controllable image editing and mask-based inpainting~\cite{meng2022sdedit,lugmayr2022repaint}, providing foundations for localized semantic image modification.

\section{Threat Model}

We consider an adversary whose objective is to infer a user's real-world location from photographs that the user has publicly shared on social or commercial platforms, such as social media posts or online listings (e.g., Instagram, Discord, Facebook, e-commerce listings). In this setting, explicit metadata such as GPS coordinates and device identifiers may be removed during upload, leaving the image content itself as the primary source of location evidence.

The adversary has no access to the user's device, private metadata, or non-visual signals. The attacker operates only on the shared image and may query a multimodal model with location-seeking prompts. The model may return a predicted place, coordinate, or textual explanation that highlights visual evidence such as signs, storefronts, road layouts, vegetation, or architectural patterns. The adversary may then verify the prediction using public mapping services or other open sources. We regard an attack as successful when the prediction identifies the correct city-scale or fine-grained administrative region, since such information can substantially narrow the user's location.

The defender may be the image owner or a hosting platform that modifies the image before publication. The defense does not require knowledge of the attacker's model architecture or parameters. Instead, the defender uses an accessible multimodal model as a proxy for cue discovery, edit-instruction generation, and candidate verification. In the transfer setting, the target models are not queried during defense generation or selection; each selected image is reused across these targets. We separately report evaluation on the proxy model itself and on held-out target models (Section~\ref{sec:experiment_setup}). The defense goal is to reduce geolocation leakage by increasing localization error or lowering administrative-level accuracy, while preserving most non-sensitive scene content through localized edits. We focus on visual inference-based geolocation attacks; non-visual leakage vectors such as IP addresses, GPS tags, account metadata, or device fingerprinting are out of scope.
\section{The GeoCloak Framework}
\subsection{Empirical Observation}
\label{sec:observation}

We noticed an empirical pattern that we repeatedly observe in multimodal geolocation reasoning. Although geolocation predictions are often supported by multiple cues jointly, model-generated explanations frequently organize these cues unevenly: a small number of high-information cues serve as primary anchors, while broader contextual evidence is used to refine or corroborate the prediction.

\noindent\textbf{(1) Hierarchical Cue Dominance.}
In many examples, multimodal models first rely on distinctive cues such as textual signage, storefront names, landmarks, or highly recognizable architectural elements. They then incorporate secondary context, such as vegetation, architectural style, road layout, or surrounding scene structure, to refine the location hypothesis. As shown in Figure~\ref{fig:attention} (top), the model-generated explanation first anchors on the ``Tiffany \& Co.'' and ``Gucci'' storefront cluster, and then uses additional scene context to support the predicted location.

\noindent\textbf{(2) Compensatory Cue Reweighting.}
Removing a salient cue does not necessarily make the model fail. When one strong cue is removed or blocked, the model may reweight remaining cues and still recover the same location. In Figure~\ref{fig:attention} (bottom left), removing the primary Tiffany \& Co. facade does not change the prediction, because the model promotes other contextual signals to maintain a coherent explanation. This behavior suggests that defenses based only on erasure or masking may be insufficient when multiple cues jointly support geolocation.

By contrast, a localized edit that introduces a plausible but misleading high-information cue can shift the model's prediction in some cases. Figure~\ref{fig:attention} (bottom right) illustrates this behavior: adding a false anchor, ``PLAZA MARIE JOSE MARTI,'' changes the model-generated explanation and shifts the predicted location to a distant region. This observation motivates cue-aware semantic editing: rather than broadly obfuscating the entire image, GeoCloak targets geolocation-relevant cues through localized edits that change the evidence available to the model.

\begin{figure*}[t]
    \centering
    \includegraphics[width=\textwidth]{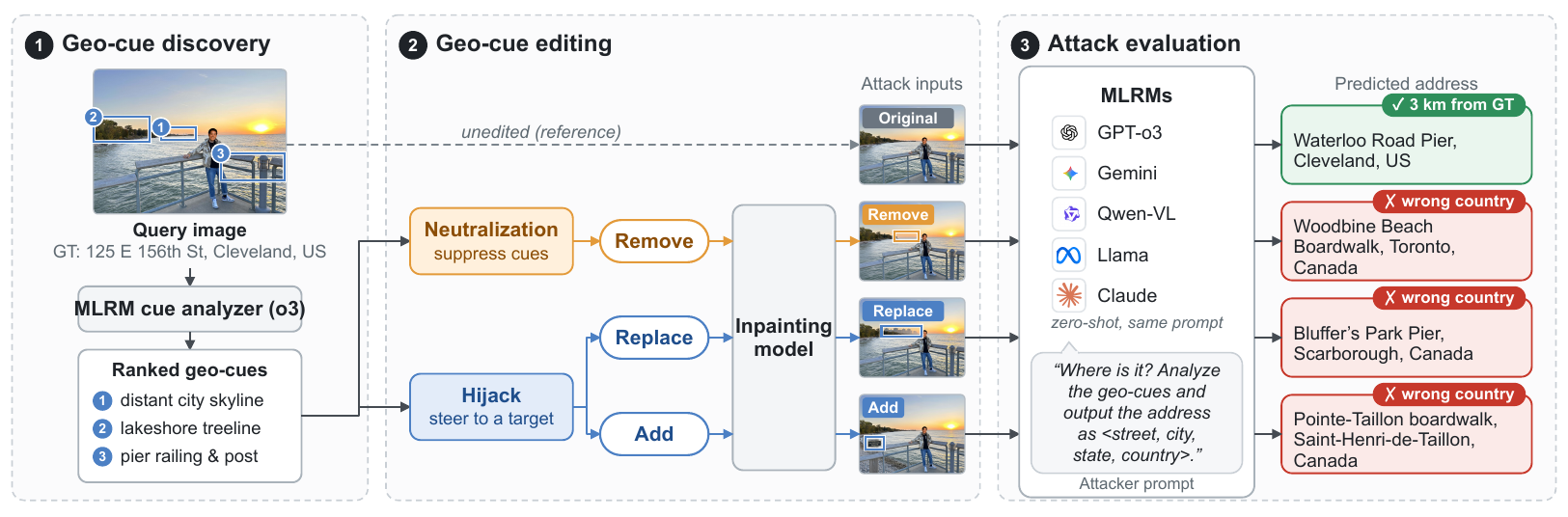}
\caption{
Overview of \textsc{GeoCloak}. A defender-side proxy model identifies and ranks geolocation-relevant cues. 
\textsc{GeoCloak} then generates localized \textsc{Remove}, \textsc{Replace}, and \textsc{Add} edits, grounds the selected region, applies inpainting, and selects the candidate with the largest great-circle displacement from the proxy's prediction on the original image.
}
    \label{fig:pipeline}\vspace{-0.2cm}
\end{figure*}

\subsection{Problem Setup and Goal}
\label{sec:problem_setup}

In this work, we study defenses against visual geolocation attacks, where an adversary seeks to infer the true location $y$ associated with a publicly shared image $I$. The adversary uses a black-box multimodal geolocation model $f$ that predicts a location $\hat{y}_f=f(I)$ by aggregating evidence from multiple observable geo-cues embedded in $I$. These cues $\{C_i\}$, such as street signs, architectural styles, road markings, storefront text, or vegetation, may jointly support the model's prediction.

Our goal is to reduce the geolocation information available to the model through localized cue-level edits. Specifically, we ask:

\textbf{\emph{Can a small localized edit $C_i \to C_i'$, applied to one or a few geolocation-relevant cues in $I$, reduce the accuracy of the model's location prediction while preserving most of the original scene content?}}

We directly leverage the MLRM's own reasoning about which visual elements support its prediction, and prompt it to propose edits that are both minimal and semantically consistent (i.e., small, localized changes that preserve the global scene semantics). In this way, realism and minimality are enforced via the model's reasoning process, without introducing external hand-crafted fidelity constraints.

The defender aims to produce an edited image $I^\star$ that reduces the target model's geolocation accuracy, evaluated against $y$, while limiting changes to selected cue regions. Candidate selection uses a defender-side proxy $g$, whose prediction may differ from that of the target $f$; the selection rule is specified in Section~\ref{subsec:verification}.

\subsection{System Overview}
\label{sec:overview}

Figure~\ref{fig:pipeline} shows the end-to-end workflow of \textsc{GeoCloak}. The pipeline has four stages:

\begin{enumerate}
    \item \textbf{Cue discovery and ranking.} 
    A proxy multimodal model analyzes the input image and identifies candidate geolocation-relevant cues, such as text, signs, landmarks, road markings, or architectural patterns. These cues are ranked by a score that balances estimated geolocation relevance and spatial footprint.

    \item \textbf{Edit strategy generation.}
    For the top-ranked cue, GeoCloak generates three localized edit strategies: \textsc{Remove}, \textsc{Replace}, and \textsc{Add}. These strategies correspond to two goals: neutralizing existing geolocation evidence or introducing plausible but misleading evidence.

    \item \textbf{Region grounding and image editing.}
    The selected cue region is localized using GroundedSAM~\cite{ren2024grounded}. The edit strategies are then realized through diffusion-based inpainting, producing candidate edited images that aim to preserve the surrounding scene content.

    \item \textbf{Verification and selection.}
    The defender-side proxy predicts a location for the original image and each candidate edit. GeoCloak selects the candidate with the largest great-circle displacement between the original and edited-image predictions, as detailed in Section~\ref{subsec:verification}.
\end{enumerate}

\subsection{System Implementation}
\label{sec:pipeline_impl}

\subsubsection{Cue Discovery \& Ranking}
\label{sec:cues}

GeoCloak first uses an accessible multimodal model as a proxy to identify visual evidence that may support geolocation inference. Given an input image $I$, the proxy model returns a set of localized candidate cues:
\begin{equation}
\mathcal{C}=\{c_i=(t_i,R_i,r_i)\}_{i=1}^{N},
\end{equation}
where $t_i$ denotes the cue type (e.g., text, sign, landmark, road marking, or architectural pattern), $R_i \subset \Omega$ denotes the corresponding image region, and $r_i\in[0,1]$ is a proxy-model estimate of the cue's relevance for geolocation. 

To encourage localized edits, GeoCloak ranks candidate cues using
\begin{equation}
\label{eq:cue_score}
\tilde{s}_i = w_s(1-v_i)+w_r r_i, 
\qquad w_s+w_r=1,
\end{equation}
where $v_i$ is the normalized area footprint of cue $c_i$. The first term favors spatially small edits, while the second favors cues estimated to be informative for geolocation. We select the top-ranked cue $c^\star$ with region $R^\star$ for editing. In all experiments, we fix $w_s=0.8$ and $w_r=0.2$ without tuning these weights on test outcomes. We evaluate this cue-selection rule against lowest-ranked and random-cue selection in Section~\ref{subsec:cue_selection_ablation}.

\subsubsection{Edit Strategy Generation}
\label{sec:strategy}

Given the selected cue $c^\star$, GeoCloak generates localized semantic edits that aim to change the geolocation evidence available to the model while preserving the surrounding scene content. We consider three edit operations: \textsc{Remove}, \textsc{Replace}, and \textsc{Add}. \textsc{Remove} weakens or removes existing geolocation evidence; \textsc{Replace} substitutes the selected cue with a less diagnostic or misleading alternative; and \textsc{Add} introduces a plausible cue that may shift the model's location hypothesis.

Because the effect of an edit is difficult to predict before generation, GeoCloak produces multiple candidate edits and evaluates them after image editing. We use prompt guidelines to encourage edits that are localized, visually integrated, and scene-consistent. Representative prompt templates, positive/negative examples, and operational guidelines are provided in Appendix~\ref{app:prompts}.

Formally, for the selected cue $(c^\star,R^\star)$, the strategy module defines
\begin{equation}
\Phi_j(I; R^\star, c^\star),\quad j \in \{\textsc{remove}, \textsc{replace}, \textsc{add}\}.
\end{equation}
where $\Phi_j$ specifies how region $R^\star$ should be edited under operation $j$.

\subsubsection{Region Grounding and Image Editing}
\label{subsec:editing}

GeoCloak localizes the selected cue region using GroundedSAM~\cite{ren2024grounded}. Given the cue description and image $I$, GroundedSAM produces a candidate mask $M^\star$ for the region to be edited. This mask constrains the edit to a localized area, reducing changes to unrelated scene content.

Each edit strategy is then realized by an inpainting model:
\begin{equation}
I'_j=\mathcal{R}(I;M^\star,\Phi_j),
\end{equation}
where $\mathcal{R}$ denotes the image editor conditioned on the input image, mask, and textual edit instruction. This produces a set of candidate edited images
\[
\mathcal{I}=\{I'_{\textsc{remove}}, I'_{\textsc{replace}}, I'_{\textsc{add}}\}.
\]

\subsubsection{Verification and Selection}
\label{subsec:verification}

GeoCloak evaluates each candidate edit using a defender-side geolocation evaluator. In our experiments, this evaluator is OpenAI o3 (\texttt{o3-2025-04-16}), the same proxy used for cue discovery and edit-instruction generation. Let $g$ denote this evaluator. For the original image and each candidate edit, we obtain predicted locations $\hat{y}=g(I)$ and $\hat{y}'_j=g(I'_j)$ when coordinate predictions are available.

We measure the displacement using the great-circle (Haversine) distance:
\begin{equation}
D_j=d(\hat{y},\hat{y}'_j),
\end{equation}
where $d(\cdot,\cdot)$ is the Haversine distance in kilometers. GeoCloak selects the candidate with the largest displacement under this proxy:
\begin{equation}
I^\star=I'_{j^\star},
\qquad
j^\star=\arg\max_{j\in\{\textsc{remove},\textsc{replace},\textsc{add}\}}D_j.
\end{equation}

This is a label-free candidate-selection heuristic: it does not use the true location $y$. Final evaluation separately measures country-, state-, city-, and street-level geolocation accuracy against ground-truth labels. %The distinction between prediction displacement and ground-truth error is discussed in the Limitations section.
Section~\ref{subsec:operation_selection_ablation} compares this selection rule with fixed \textsc{Remove}, \textsc{Replace}, and \textsc{Add} choices over the same pre-generated candidates.

\section{Experiment}

\begin{table}[t]
\centering
\scriptsize
\setlength{\tabcolsep}{2.2pt}
\renewcommand{\arraystretch}{0.95}
\begin{tabular*}{\columnwidth}{@{\extracolsep{\fill}}llrrrr@{}}
\toprule
\textbf{Model} & \textbf{Method} 
& \textbf{Ctry.} & \textbf{State} & \textbf{City} & \textbf{Street} \\
& & \textbf{Acc.} & \textbf{Acc.} & \textbf{Acc.} & \textbf{Acc.} \\
\midrule

\textbf{o3}$\dagger$ 
& None  & 97.70 & 72.41 & 28.64 & 13.37 \\
& Gauss & 97.12 & 72.98 & 32.57 & 12.28 \\
& M-Att & 91.90 & 47.39 &  7.54 &  2.51 \\
& Ours   & \textbf{72.18} & \textbf{39.64} & \textbf{9.35} & \textbf{3.14} \\
\midrule

\textbf{GPT-5} 
& None  & 98.84 & 91.32 & 32.16 & 13.45 \\
& Gauss & 99.42 & 90.75 & 30.63 & 16.76 \\
& M-Att & 91.90 & 60.11 &  8.80 &  4.14 \\
& Ours   & \textbf{83.43} & \textbf{60.72} & \textbf{9.46} & \textbf{3.14} \\
\midrule

\textbf{GPT-4o} 
& None  & 81.60 & 30.45 & 14.86 &  2.77 \\
& Gauss & 83.42 & 22.28 &  8.52 &  3.37 \\
& M-Att & 60.69 &  5.78 &  1.88 &  0.00 \\
& Ours   & \textbf{46.74} & \textbf{2.95} & \textbf{1.76} & \textbf{0.59} \\
\midrule

\textbf{Gemini} 
& None  & 98.85 & 91.37 & 44.57 & 12.06 \\
& Gauss & 97.14 & 89.14 & 39.20 & 17.44 \\
& M-Att & 83.51 & 64.16 & 10.34 &  2.75 \\
& Ours   & \textbf{65.68} & \textbf{56.21} & \textbf{11.76} & \textbf{2.95} \\
\midrule

\textbf{Opus-4} 
& None  & 89.08 & 75.86 & 21.14 & 10.19 \\
& Gauss & 56.57 & 49.71 & 11.93 &  7.92 \\
& M-Att & 72.25 & 32.94 &  4.00 &  2.40 \\
& Ours   & \textbf{39.05} & \textbf{31.36} & \textbf{5.29} & \textbf{2.36} \\
\midrule

\textbf{Llama-4} 
& None  & 87.93 & 71.83 & 16.57 &  3.87 \\
& Gauss & 89.08 & 62.64 & 14.29 &  4.43 \\
& M-Att & 50.86 & 29.47 &  2.27 &  0.00 \\
& Ours   & \textbf{49.70} & \textbf{21.30} & \textbf{2.94} & \textbf{0.59} \\
\midrule

\textbf{Sonnet-4} 
& None  & 48.85 & 41.37 & 15.43 & 12.64 \\
& Gauss & 30.85 & 28.00 & 10.80 & 10.71 \\
& M-Att & 19.88 & 10.52 & N/A   & N/A \\
& Ours   & \textbf{14.20} & \textbf{10.65} & \textbf{2.94} & \textbf{1.18} \\
\midrule

\textbf{Qwen} 
& None  & 47.70 & 44.25 & 12.57 & 7.05 \\
& Gauss & 42.52 & 40.22 & 12.00 & 9.21 \\
& M-Att & 19.58 &  5.84 & N/A   & N/A \\
& Ours   & \textbf{10.05} & \textbf{8.28} & \textbf{1.18} & \textbf{0.00} \\
\bottomrule
\end{tabular*}
\caption{Top-1 geolocation accuracy (\%) on the Doxing dataset. Lower is better. None, Gauss, M-Att, and Ours denote no defense, Gaussian noise, M-Attack-U, and \textsc{GeoCloak}, respectively. For \textsc{GeoCloak}, o3 ($\dagger$) also serves as the defender-side proxy; the other seven targets are held out during defense generation and selection. The same selected images are evaluated across targets. N/A indicates unparseable city- or street-level predictions.}
\label{tab:doxing_strong}
\end{table}

\subsection{Experiment Setup}
\label{sec:experiment_setup}
We evaluate \textsc{GeoCloak} under the iterative, query-based geolocation attack pipeline of Luo et al.~\cite{luo2025doxing}.
The main Doxing comparison evaluates no defense, Gaussian noise, M-Attack-U, and \textsc{GeoCloak} against eight multimodal large reasoning models (MLRMs): OpenAI \emph{GPT-5}, OpenAI \emph{o3}, OpenAI \emph{GPT-4o}, Google \emph{Gemini~2.5~Pro}, Anthropic \emph{Claude~Opus-4}, Anthropic \emph{Claude Sonnet~4}, Meta \emph{Llama-4~Maverick}, and \emph{Qwen2.5-VL}.

\noindent\textbf{Defender-side proxy.}
We use OpenAI o3 (\texttt{o3-2025-04-16}) for cue discovery, edit-instruction generation, and candidate verification. GroundedSAM localizes the selected cue, and the inpainting model renders the \textsc{Remove}, \textsc{Replace}, and \textsc{Add} candidates from the generated instructions.

\noindent\textbf{Cross-model protocol.}
For each input, we select one final protected image and reuse it across the target models evaluated on that dataset. None of the seven non-o3 targets is queried during cue discovery, edit generation, verification, or candidate selection. We report \textsc{GeoCloak}'s results on these held-out models as cross-model transfer, and distinguish them from o3-on-o3 evaluation, where the target also serves as the defense proxy.

% \noindent\textbf{Editing configuration and candidate count.}
% For each input, we generate one candidate for each of \textsc{Remove}, \textsc{Replace}, and \textsc{Add}, giving three candidates before verification-based selection. The reported local timing measurement is approximately 25 seconds for one 50-step inpainting output on a single NVIDIA A100 80GB GPU. This is not the end-to-end latency of producing a protected image. Stage-wise query accounting is given in Appendix~\ref{app:cost}, with configuration fields and a per-image execution-record template in Appendices~\ref{app:repro_config} and~\ref{app:execution_record}.

\noindent\textbf{Editing configuration and cost.}
For each input, we generate one candidate for each of \textsc{Remove}, \textsc{Replace}, and \textsc{Add} using FLUX.1 Fill-dev with 50 sampling steps, giving three candidates before verification-based selection. Protecting one image requires one GroundedSAM run, three inpainting calls (approximately 25\,s each on a single NVIDIA A100 80GB GPU), and five to six proxy queries: one for cue discovery, one for edit-instruction generation, three for candidate verification, and one for the original image unless an earlier prediction is reused. These counts exclude retries and downstream target-model evaluation.

\subsection{Dataset}
Our evaluation uses two complementary datasets that capture distinct aspects of geolocation and privacy risk:

\begin{itemize}
\item \textbf{Doxing Dataset}~\cite{luo2025doxing}: 172 social-media-style images (Level-1 Risk and Mirror subsets) emphasizing fine-grained, high-information cues such as storefronts, signage, or architectural styles that often enable real-world doxing.
\item \textbf{FairLocator}~\cite{huang2025vlms}: 600 globally distributed street-view images from 60 cities, emphasizing generalization to diverse urban contexts with low-saliency, contextual cues.
\end{itemize}

These two datasets complement each other:
Doxing focuses on user-generated, privacy-sensitive imagery rich in localized identifiers and challenging viewing conditions, while FairLocator represents standardized street scenes where location inference depends on broad contextual understanding.
Together, they enable a comprehensive evaluation of both targeted (fine-grained) and generalized (global) localization performance. Detail samples are shown in Fig.~\ref{fig:dataset_comparison}.

\begin{table}[t]
\centering
\small
\setlength{\tabcolsep}{3.5pt}
\begin{tabular}{llccc}
\hline
Model & Method & Ctry. & City & Street \\
      &        & Acc. & Acc. & Acc. \\
\hline
OpenAI o3 & No defense & 74.08 & 58.02 & 3.01 \\
          & GeoCloak   & 34.44 & 24.08 & 0.66 \\
\hline
GPT-4o    & No defense & 34.49 & 26.87 & 1.49 \\
          & GeoCloak   & 12.84 &  2.11 & 0.34 \\
\hline
Gemini 2.5 Pro & No defense & 86.73 & 67.33 & 4.81 \\
               & GeoCloak   & 44.41 & 26.38 & 0.67 \\
\hline
\end{tabular}
\caption{Top-1 geolocation accuracy (\%) on the FairLocator dataset. State-level labels are unavailable for FairLocator, so State accuracy is omitted. Lower accuracy indicates stronger privacy protection.}\vspace{-0.2in}
\label{tab:breadth_summary}
\end{table}

% \begin{figure*}[t]
%     \centering
%     \includegraphics[width=\linewidth]{fig/figure4new.pdf}
%     \caption{Perceptual realism comparison. M-Attack first downscales image and introduces obvious, severe pixel distortions even at thumbnail scale, while \textsc{GeoCloak} preserves natural, high-resolution appearance.
%     }
%     \label{fig:perceptual_gallery}
% \end{figure*}

\subsection{Defense Baselines} 
Prior work has not directly studied our defense setting, namely privacy-preserving geolocation protection via localized semantic edits. 
Therefore, we select comparators based on the closest functional objective, which is to induce mislocalization through image-level modifications in a black-box setting. 
Beyond \textsc{No Defense} and additive Gaussian noise ($\sigma=0.1$), we adopt M-Attack~\cite{li2025frustratingly} as a strong state-of-the-art perturbation baseline for misleading multimodal models. 
While M-Attack is originally formulated as a \emph{targeted} attack that steers predictions toward a specified destination, our goal is purely to cause mislocalization without prescribing where the prediction should move. 
We thus use its \emph{untargeted} variant, denoted as M-Attack-U ($\epsilon=16/255$). 

\noindent\textbf{Localized baselines.}
We additionally compare \textsc{GeoCloak} with \textsc{Local Mask} and \textsc{Local Blur} on Doxing using o3 as the target model. Both baselines use the same top-ranked cue and GroundedSAM mask as \textsc{GeoCloak}. \textsc{Local Mask} covers the masked region with a solid white patch, whereas \textsc{Local Blur} applies Gaussian blurring ($\sigma=0.1$) within that region. Both localized baselines leave pixels outside the mask unchanged.

\subsection{Evaluation Metrics} 
We evaluate geolocation performance following ~\cite{luo2025doxing}, using hierarchical Top-1 accuracy at four administrative levels:
\emph{Country}, \emph{State}, \emph{City}, and \emph{Street}.
At each level, a prediction is counted as correct if its administrative region matches the ground truth.
Country, State, and City correctness follow standard geopolitical boundaries, while Street-level correctness requires a match in the street address or the corresponding block-level identifier. 

\section{Results and Analysis}
\label{sec:results}

\begin{table}[t]
\centering
\small
\setlength{\tabcolsep}{6pt}
\renewcommand{\arraystretch}{1.12}
\begin{adjustbox}{max width=0.48\textwidth}
\begin{tabular}{lccc}
\toprule
\textbf{Method} 
& \textbf{NIQE} $\downarrow$ 
& \textbf{BRISQUE} $\downarrow$ 
& \textbf{PIQE} $\downarrow$ \\
\midrule
No Defense (Original) 
& 3.34 [2.75, 4.66] 
& 31.10 [22.66, 45.06] 
& 40.32 [27.29, 56.98] \\
\midrule
GeoCloak-Add     
& 2.70 [1.90, 5.89] 
& 22.54 [13.04, 61.43] 
& 42.01 [32.48, 59.80] \\
GeoCloak-Remove  
& 2.56 [1.90, 5.09] 
& 21.57 [12.64, 46.88] 
& 40.21 [30.17, 53.73] \\
GeoCloak-Replace 
& 3.39 [2.78, 5.21] 
& 31.88 [23.47, 47.84] 
& 39.61 [26.95, 56.62] \\
\midrule
M-Attack-U ($\epsilon{=}16/255$) 
& 6.52 [5.17, 9.01] 
& 33.92 [25.09, 48.44] 
& 40.66 [34.01, 48.76] \\
Gaussian Noise ($\sigma{=}0.1$) 
& 8.61 [6.47, 11.56] 
& 58.13 [54.29, 61.37] 
& 63.15 [58.10, 67.01] \\
\bottomrule
\end{tabular}
\end{adjustbox}
\caption{\textbf{Automatic perceptual quality under different defenses.}
Values are reported as median [p10, p90]; lower is better.
\textsc{GeoCloak} keeps no-reference quality scores closer to the original images than pixel-level perturbation baselines.}\vspace{-0.2in}
\label{tab:perceptual_quality}
\end{table}

\begin{figure*}[t]
    \centering
    \includegraphics[width=\linewidth]{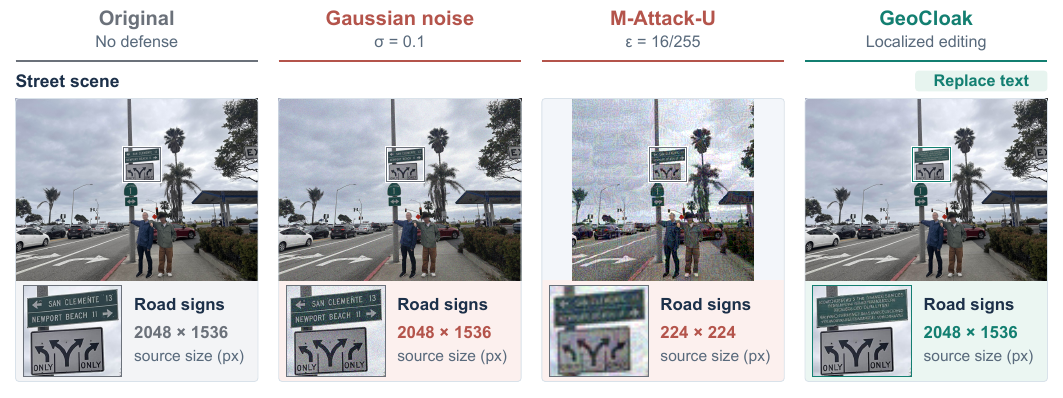}
\caption{Visual comparison of different defense methods on a representative street scene. Gaussian noise and M-Attack-U apply image-wide perturbations, while GeoCloak modifies a localized road sign and preserves most of the surrounding scene content.}
    \label{fig:perceptual_gallery}
\end{figure*}

\subsection{Geolocation Protection Results}
\label{sec:main_results}

We report Doxing results in Table~\ref{tab:doxing_strong} and FairLocator results in Table~\ref{tab:breadth_summary}. 
Across evaluated models, \textsc{GeoCloak} reduces Top-1 geolocation accuracy relative to no defense, especially at City and Street levels. 
On Doxing, City/Street accuracy drops from 28.64\%/13.37\% to 9.35\%/3.14\% on OpenAI o3, from 14.86\%/2.77\% to 1.76\%/0.59\% on GPT-4o, and from 44.57\%/12.06\% to 11.76\%/2.95\% on Gemini 2.5 Pro. 

Compared with perturbation baselines, \textsc{GeoCloak} provides a different privacy--utility trade-off. 
Gaussian noise often leaves semantic geo-cues recognizable, while M-Attack-U can reduce accuracy by making images difficult to parse; we mark unparseable City/Street outputs as N/A. 
In contrast, \textsc{GeoCloak} modifies selected semantic cues rather than globally corrupting the image.

FairLocator shows a similar trend in broader street-view scenes. 
\textsc{GeoCloak} reduces City accuracy from 58.02\% to 24.08\% on OpenAI o3, from 26.87\% to 2.11\% on GPT-4o, and from 67.33\% to 26.38\% on Gemini 2.5 Pro. These results further demonstrate GeoCloak's effectiveness in reducing geolocation accuracy across diverse street-view scenes.

\noindent\textbf{Localized-baseline comparison.}
Table~\ref{tab:localized_baselines} compares the localized baselines on Doxing with o3 as both the target model and \textsc{GeoCloak}'s proxy. With the selected cue and grounded region held fixed, full \textsc{GeoCloak} yields lower geolocation accuracy than \textsc{Local Mask} and \textsc{Local Blur} at all four administrative levels. City accuracy is 25.41\% for Mask, 23.96\% for Blur, and 9.35\% for \textsc{GeoCloak}; Street accuracy is 9.82\%, 10.12\%, and 3.14\%, respectively.

Full \textsc{GeoCloak} generates three semantic-edit candidates and applies verification-based selection, while each localized baseline applies a single fixed transformation. Unlike the single-transformation localized baselines, GeoCloak generates three semantic-edit candidates and selects the final output through proxy-based verification. The comparison therefore evaluates the effectiveness of the complete defense pipeline under the o3-based setting.

\begin{table}[t]
\centering
\small
\setlength{\tabcolsep}{3pt}
\begin{tabular*}{\columnwidth}{@{\extracolsep{\fill}}lrrrr@{}}
\toprule
Method & Country & State & City & Street \\
\midrule
No Defense & 97.70 & 72.41 & 28.64 & 13.37 \\
Local Mask & 95.52 & 66.96 & 25.41 & 9.82 \\
Local Blur & 93.15 & 64.11 & 23.96 & 10.12 \\
Full \textsc{GeoCloak} & \textbf{72.18} & \textbf{39.64} & \textbf{9.35} & \textbf{3.14} \\
\bottomrule
\end{tabular*}
\caption{Localized-baseline comparison on Doxing using o3 as the target and \textsc{GeoCloak}'s proxy. Values are Top-1 geolocation accuracy (\%); lower is better. Mask, Blur, and \textsc{GeoCloak} use the same top-ranked cue and GroundedSAM mask.}\vspace{-0.2in}
\label{tab:localized_baselines}
\end{table}

\subsection{Perceptual Quality and Visual Resolution}
\label{sec:perceptual_quality}

Beyond reducing geolocation accuracy, a useful privacy defense should preserve the visual utility of the shared image. 
We evaluate image quality using three no-reference perceptual-quality metrics: NIQE~\cite{mittal2013niqe}, BRISQUE~\cite{mittal2012brisque}, and 
PIQE~\cite{venkatanath2015pique}, where lower scores indicate better perceptual quality.
These metrics are appropriate for our setting because users typically share a single edited image, and viewers judge whether that image appears visually natural without comparing it to the original.

Table~\ref{tab:perceptual_quality} reports median scores with 10th/90th percentile ranges. 
Overall, \textsc{GeoCloak} maintains automatic perceptual-quality scores closer to those of the original images than pixel-level perturbation baselines. 
For example, the original images have a median NIQE of 3.34, while \textsc{GeoCloak}-Remove and \textsc{GeoCloak}-Add obtain median NIQE scores of 2.56 and 2.70. 
By contrast, M-Attack-U increases median NIQE to 6.52, and Gaussian noise increases it to 8.61. 
A similar pattern appears for BRISQUE, where \textsc{GeoCloak} variants remain closer to the original distribution than Gaussian noise.

Figure 4 illustrates the visual differences among the defense methods on a representative street scene. While Gaussian noise introduces relatively subtle changes and M-Attack-U produces more noticeable image-wide artifacts, GeoCloak modifies the selected road sign while preserving most surrounding content. This highlights the potential benefit of localized semantic editing for maintaining the visual usability of protected images.

% To complement automatic metrics, we conduct a preliminary human naturalness study with 20 participants.
% Participants are shown a randomized mixture of original images, Gaussian-noise images, M-Attack-U images, and \textsc{GeoCloak}-edited images, without being told which condition each image belongs to.
% For each image, they judge whether it looks visually natural as a photograph, selecting ``natural'' or ``unnatural.'' The naturalness rate is the percentage of judgments labeled ``natural'' for each condition.
% As shown in Table~\ref{tab:human_naturalness}, the rates are 98.1\% for original images, 30.6\% for Gaussian noise, 2.4\% for M-Attack-U, and 85.4\% for \textsc{GeoCloak}.
% Thus, \textsc{GeoCloak} receives a higher proportion of naturalness judgments than the two perturbation baselines in this study, while remaining below the original images. Participant details and the full instruction are provided in Appendix~\ref{app:human_study}.

\noindent\textbf{Human naturalness study.}
We also conduct a preliminary human study with 20 voluntary, unpaid participants recruited through convenience sampling; no demographic or identifying information was collected.
Participants view a randomized mix of original, Gaussian-noise, M-Attack-U, and \textsc{GeoCloak} images without knowing each image's condition, and label each one as ``natural'' or ``unnatural'' as a photograph, judging visible artifacts and physical plausibility rather than geographic correctness.
The resulting naturalness rates (share of ``natural'' labels) are 98.1\% for original images, 85.4\% for \textsc{GeoCloak}, 30.6\% for Gaussian noise, and 2.4\% for M-Attack-U: \textsc{GeoCloak} is judged natural far more often than either perturbation baseline, though still below the unmodified images.
%Appendix~\ref{app:backbone} asks whether the effect depends on a particular image editor, and reports qualitative comparisons across different inpainting backbones. 
%In Appendix~\ref{app:end2end}, directly prompting a general-purpose image-editing model(Gemini 3 Pro Image) to ``hide the location'' tended to alter broad, unrelated regions while leaving residual cues, which motivates explicit cue selection and grounding before editing.

% 这里要看留什么，这里的几个都是对应了一部分的附录实验

% Appendix~\ref{app:end2end} presents a qualitative pilot using an end-to-end ``hide the location'' prompt; it is not a quantitative comparison under matched candidate or query budgets.

% \begin{table}[t]
% \centering
% \small
% \begin{tabular}{lc}
% \toprule
% Condition & Naturalness Rate (\%) \\
% \midrule
% Original images & 98.1 \\
% Gaussian noise & 30.6 \\
% M-Attack-U & 2.4 \\
% \textsc{GeoCloak} & 85.4 \\
% \bottomrule
% \end{tabular}
% \caption{Preliminary human naturalness study with 20 participants. For each condition, the naturalness rate is the percentage of judgments labeled ``natural'' in a binary natural/unnatural task.}
% \label{tab:human_naturalness}
% \end{table}

% \label{app:backbone}

\noindent\textbf{Image Editor Sensitivity.} We qualitatively examine the sensitivity of \textsc{GeoCloak} to the choice of image editor. In addition to FLUX.1 Fill-dev~\cite{bfl2024fluxfill}, used in our main experiments, we evaluate Gemini 2.5 Flash Image (Nano Banana)~\cite{google2025nanobanana} as an alternative editor. We keep the input image, selected cue, and edit instruction fixed across editors.

\begin{figure}[t]
  \centering
  \includegraphics[width=0.8\linewidth]{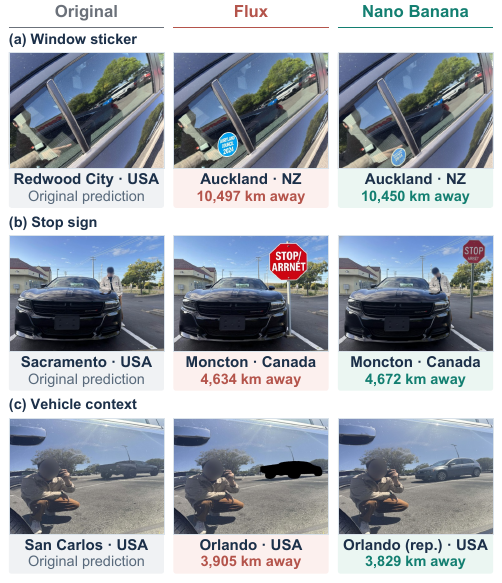}
  \caption{Qualitative comparison of FLUX.1 Fill-dev and Gemini 2.5 Flash Image (Nano Banana) under the same cue selection and edit instruction, together with downstream geolocation predictions.}
  \label{fig:quality_backbone}
\end{figure}

Figure~\ref{fig:quality_backbone} presents representative edited outputs and their downstream geolocation predictions. These examples illustrate how different image editors can produce varying visual modifications and geolocation outcomes under the same editing objective.

\noindent\textbf{Direct End-to-End Editing Pilot.}
\label{app:end2end}
We also investigate a simpler alternative: directly prompting Gemini 3 Pro Image (Nano Banana Pro)~\cite{google2025nanobananapro} to ``hide the location'' while preserving visual realism, without explicit cue ranking or region grounding.

\begin{figure}[t]
  \centering
  \includegraphics[width=0.8\linewidth]{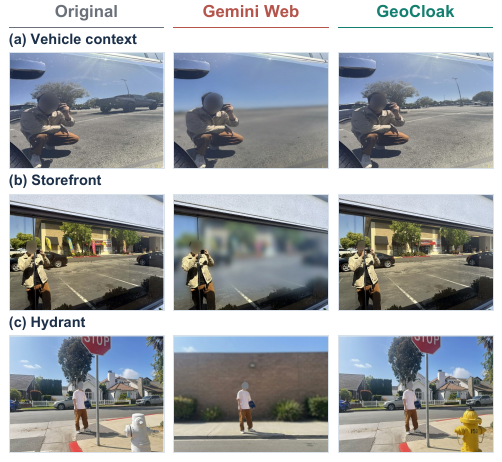}
  \caption{Qualitative examples of direct end-to-end location-hiding edits using Gemini 3 Pro Image, illustrating changes to unrelated scene content and residual geolocation cues.}\vspace{-0.1in}
  \label{fig:gemini}
\end{figure}

As illustrated in Figure~\ref{fig:gemini}, direct editing can introduce broad modifications to unrelated scene content while leaving some geolocation cues intact. These observations motivate \textsc{GeoCloak}'s explicit cue selection and localized editing. This pilot provides a qualitative illustration of potential failure modes rather than a controlled quantitative comparison.

Together with the geolocation results, these evaluations provide preliminary evidence about privacy protection, perceptual quality, and visual naturalness.

\subsection{Ablation Studies}
\label{sec:ablations}

\subsubsection{Cue Selection}
\label{subsec:cue_selection_ablation}
On Doxing, we compare selecting the highest-ranked cue, the lowest-ranked cue (reported as \textsc{Low-ranked}), and a randomly selected cue from the discovered cue list. The ranking follows the combined area-and-relevance score in Eq.~\ref{eq:cue_score}. We use o3 as both the target model and the defender-side proxy, and keep the editing and verification procedures unchanged across conditions. Random-cue selection is repeated three times, and we report the average result for this condition.

\begin{table}[t]
\centering
\small
\setlength{\tabcolsep}{3pt}
\begin{tabular*}{\columnwidth}{@{\extracolsep{\fill}}lrrrr@{}}
\toprule
Cue selection & Country & State & City & Street \\
\midrule
Low-ranked & 96.51 & 65.74 & 20.63 & 10.59 \\
Random & 87.32 & 57.41 & 16.31 & 7.52 \\
Top-ranked & \textbf{72.18} & \textbf{39.64} & \textbf{9.35} & \textbf{3.14} \\
\bottomrule
\end{tabular*}
\caption{Cue-selection ablation on Doxing using o3 as both the target and defender-side proxy. Values are Top-1 geolocation accuracy (\%); lower is better. Low-ranked selects the lowest-ranked discovered cue. Random reports the average over three runs; editing and verification follow the same procedure across conditions.}\vspace{-0.2in}
\label{tab:cue_selection_ablation}
\end{table}

Table~\ref{tab:cue_selection_ablation} shows that selecting the top-ranked cue yields the lowest geolocation accuracy at all four administrative levels. City accuracy is 20.63\% with the lowest-ranked cue, 16.31\% with a random cue, and 9.35\% with the top-ranked cue; Street accuracy is 10.59\%, 7.52\%, and 3.14\%, respectively. These results demonstrate that the proposed cue-ranking strategy consistently outperforms random and low-ranked cue selection across all four administrative levels in the o3-based evaluation.

\subsubsection{Edit Operations and Candidate Selection}
\label{subsec:operation_selection_ablation}
On Doxing with o3 as both the target model and the defender-side proxy, we compare fixed \textsc{Remove}-only, \textsc{Replace}-only, and \textsc{Add}-only choices with verification-based selection. For each input, all four policies reuse the same three pre-generated candidates, one per edit operation. Each fixed policy outputs the candidate corresponding to its designated operation; verification-based selection chooses the candidate with the largest proxy-prediction displacement, following Section~\ref{subsec:verification}. This design holds the candidate set fixed while varying the output-selection rule.

\begin{table}[t]
\centering
\small
\setlength{\tabcolsep}{3pt}
\begin{tabular*}{\columnwidth}{@{\extracolsep{\fill}}lrrrr@{}}
\toprule
Selection rule & Country & State & City & Street \\
\midrule
Remove-only & 95.56 & 68.45 & 22.13 & 11.86 \\
Replace-only & 75.86 & 46.57 & 15.39 & 9.25 \\
Add-only & \textbf{71.42} & 42.67 & 13.24 & 4.86 \\
Verification & 72.18 & \textbf{39.64} & \textbf{9.35} & \textbf{3.14} \\
\bottomrule
\end{tabular*}
\caption{Edit-operation and candidate-selection ablation on Doxing using o3 as both the target and defender-side proxy. Values are Top-1 geolocation accuracy (\%); lower is better. All four policies reuse the same three pre-generated candidates per input. Fixed policies select the candidate for the named operation; Verification selects by proxy-prediction displacement.}\vspace{-0.2in}
\label{tab:operation_selection_ablation}
\end{table}

Table~\ref{tab:operation_selection_ablation} shows that verification-based selection achieves the lowest State, City, and Street accuracy (39.64\%, 9.35\%, and 3.14\%, respectively). \textsc{Add}-only achieves the lowest Country accuracy (71.42\%, compared with 72.18\% for verification). Relative to \textsc{Add}-only, the best fixed-operation policy at each level in this comparison, verification reduces City and Street accuracy by 3.89 and 1.72 percentage points, respectively. These results support the contribution of verification-based candidate selection at the state, city, and street levels in this configuration, while showing that it is not the best policy at every level. These results demonstrate that verification-based candidate selection improves fine-grained geolocation protection over fixed editing operations, achieving the lowest State, City, and Street accuracy while remaining competitive at the Country level.

\section{Conclusion}
We presented \textsc{GeoCloak}, a cue-aware perceptual editing framework for reducing geolocation inference from user-shared images. By modifying localized semantic cues rather than applying global pixel perturbations, \textsc{GeoCloak} reduces administrative-level geolocation accuracy across multiple multimodal models while maintaining automatic perceptual-quality scores closer to those of the original images than perturbation baselines. These results suggest that cue-level semantic editing is a promising direction for reasoning-based visual privacy. %Future work should address human-perceived semantic plausibility, adaptive attackers, platform post-processing, and more efficient deployment.

\section*{Ethical Considerations}
GeoCloak is intended to help users reduce involuntary location leakage, but semantic editing is dual-use and could be misused to manipulate visual context; we evaluate it only in controlled research settings and do not redistribute privacy-sensitive images beyond the original datasets' access conditions.

\section*{GenAI Usage Disclosure}
Generative AI tools were used to assist with manuscript preparation, including light language editing of author-written text and generating base graphics for some figures, which were subsequently refined and assembled by the authors. These tools were not used to design the study, generate or analyze experimental data, or write the technical content. All scientific claims and results were produced and verified by the authors. The multimodal language models and image editing models described in the paper are components of the studied system and its evaluation, rather than manuscript preparation tools, and their use is documented in the main text.

\bibliographystyle{ACM-Reference-Format}
\bibliography{custom}

\clearpage
\appendix

\section{Prompt Templates and Operational Guidelines}
\label{app:prompts}

This section provides representative prompt templates and operational guidelines used by \textsc{GeoCloak}. These templates summarize the intended inputs, output fields, and editing guidelines; they are not a verbatim archive of the API requests used in each experiment. We refer to model outputs as generated explanations or reasoning narratives, rather than assuming access to hidden model chain-of-thought.

\subsection{Geo-Cue Analysis Prompt}

The defender-side proxy model first analyzes the input image and identifies geolocation-relevant cues. The prompt asks the model to list visible cues, categorize them by type, estimate their relevance for geolocation, and propose candidate cues for localized editing.

\begin{lstlisting}[basicstyle=\ttfamily\scriptsize,breaklines=true,frame=single]
You are a geolocation analyst. Analyze the image and identify visual evidence that may support a location prediction.

Output JSON with:
1. identified_geo_cues: a list of cues with type, region description, estimated relevance score, and explanation.
2. overall_scene_summary: a brief description of the scene.
3. likely_location_hypotheses: top candidate regions or locations.
4. generated_explanation: a concise explanation of how the cues support the location.
5. candidate_edit_cues: cues that could be removed, replaced, or modified to reduce geolocation leakage.
\end{lstlisting}

\subsection{Edit Operation Guidelines}

Given a selected cue, \textsc{GeoCloak} generates candidate edit instructions for three operations: \textsc{Remove}, \textsc{Replace}, and \textsc{Add}. The goal is not to globally corrupt the image, but to modify localized geolocation evidence while preserving surrounding scene content.

\paragraph{\textsc{Remove}.}
The \textsc{Remove} operation weakens or removes an existing geolocation cue.

\begin{lstlisting}[basicstyle=\ttfamily\scriptsize,breaklines=true,frame=single]
Remove operation:
Goal: weaken or remove geolocation evidence while preserving surrounding scene content.

Guidelines:
- Prefer geographically specific text, signs, symbols, logos, or local identifiers.
- Avoid removing large unrelated regions.
- Preserve lighting, texture, and background consistency.
- The edited region should blend into the surrounding scene.
\end{lstlisting}

\paragraph{\textsc{Replace}.}
The \textsc{Replace} operation substitutes a selected cue with a less diagnostic or misleading alternative.

\begin{lstlisting}[basicstyle=\ttfamily\scriptsize,breaklines=true,frame=single]
Replace operation:
Goal: replace a selected geolocation cue with a less diagnostic or misleading alternative.

Guidelines:
- Prefer subtle replacements over obvious contradictions.
- Modify small components when possible, such as text, symbols, or signage.
- Maintain contextual harmony with weathering, lighting, perspective, and material.
- Avoid changes that make the image globally inconsistent.
\end{lstlisting}

\paragraph{\textsc{Add}.}
The \textsc{Add} operation introduces a plausible cue that may shift the model's location hypothesis.

\begin{lstlisting}[basicstyle=\ttfamily\scriptsize,breaklines=true,frame=single]
Add operation:
Goal: introduce a plausible cue that may shift the model's location hypothesis.

Guidelines:
- Integrate the cue into an existing surface or plausible object when possible.
- Add infrastructure or signs only when contextually plausible.
- Prefer regional ambiguity over extreme visual contradiction.
- Specify placement, scale, lighting, perspective, and integration details.
\end{lstlisting}

\subsection{Positive and Negative Examples}

We use lightweight positive and negative examples to guide the edit operation. These examples are not hard constraints; they are intended to discourage overly broad or visually inconsistent edits.

\begin{itemize}
    \item \textbf{Positive \textsc{Remove}:} remove a city-specific transit logo, storefront name, or local road sign while reconstructing the surrounding surface.
    \item \textbf{Negative \textsc{Remove}:} remove a large unrelated building, road, or sky region that is not the selected cue.
    \item \textbf{Positive \textsc{Replace}:} replace a highly diagnostic sign with a less specific but context-compatible sign.
    \item \textbf{Negative \textsc{Replace}:} replace a cue with an obviously incompatible object, such as adding a sign from a very different country in an otherwise unchanged scene.
    \item \textbf{Positive \textsc{Add}:} add a small, plausible poster, plaque, or street sign integrated into an existing wall or pole.
    \item \textbf{Negative \textsc{Add}:} add a floating or poorly scaled object that does not match lighting, perspective, or scene context.
\end{itemize}

\section{Dataset Details}
\label{sec:app-datasets}
\begin{figure}[t]
  \centering
  \setlength{\tabcolsep}{0.5pt}
  \begin{tabular}{cccc}
    \multicolumn{2}{c}{\footnotesize Doxing} & \multicolumn{2}{c}{\footnotesize FairLocator} \\
    \includegraphics[height=0.205\linewidth]{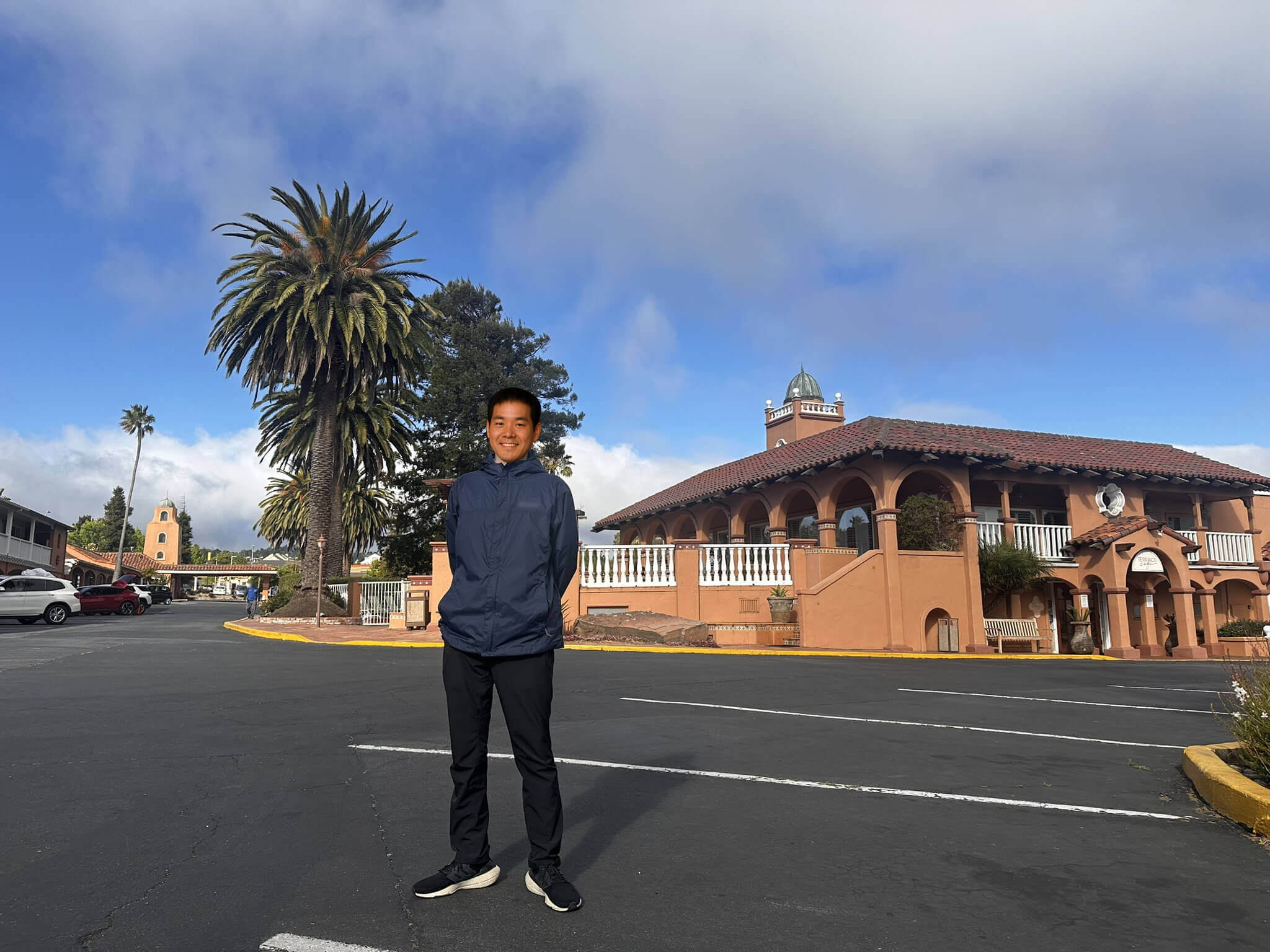} &
    \includegraphics[height=0.205\linewidth]{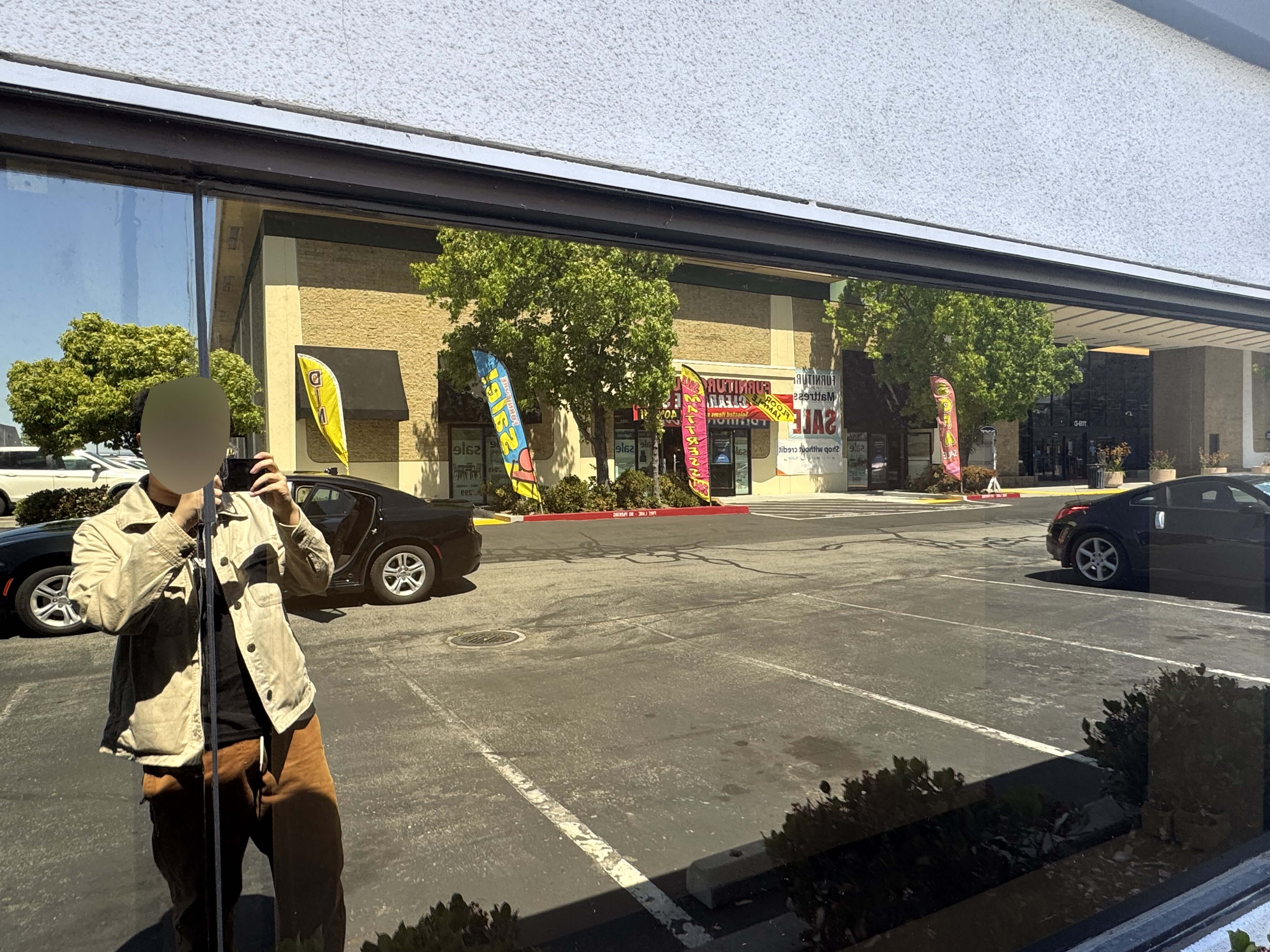} &
    \includegraphics[height=0.205\linewidth]{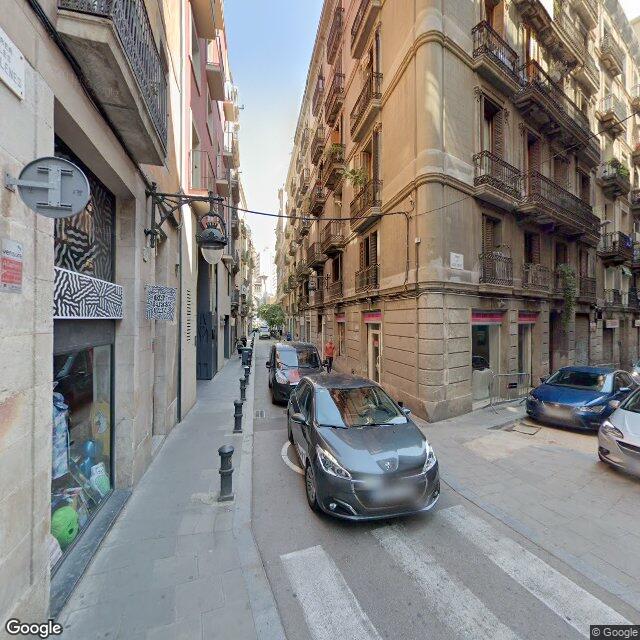} &
    \includegraphics[height=0.205\linewidth]{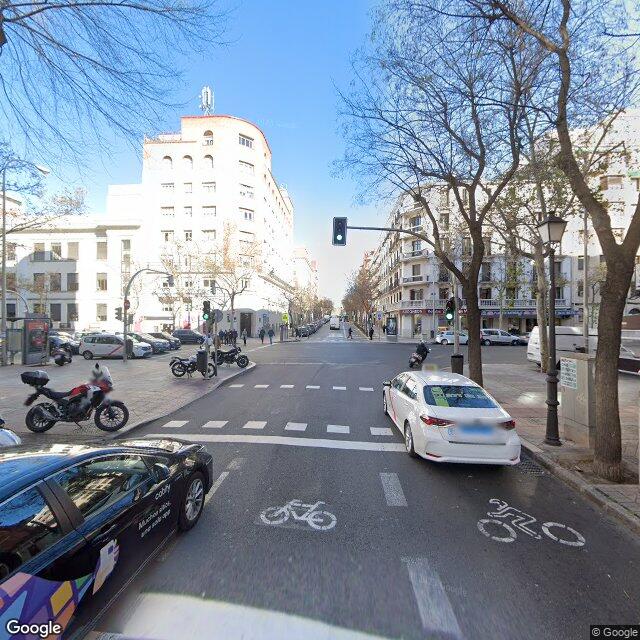}
  \end{tabular}
  \caption{Example images. Doxing (left) contains specific, localized cues; FairLocator (right) contains generic street views for broad geographic evaluation.}
  \label{fig:dataset_comparison}
\end{figure}

Our evaluation is conducted on two datasets, examples are shown in Fig.~\ref{fig:dataset_comparison}, each chosen for its unique characteristics relevant to real-world geolocation and privacy challenges.

% \subsection{The Doxing Dataset}
% % 把 “The Doxing Dataset” 相关描述、划分、预处理等内容放到这里。
% \label{sec:appendix_dataset}

% This subset is derived from the work of ~\cite{luo2025doxing} in \textit{``Doxing via the Lens: Revealing Location-related Privacy Leakage on Multi-modal Large Reasoning Models''}. The original dataset is designed to simulate real-world doxing scenarios where images, often posted on social media, contain subtle but critical clues that can lead to privacy leakage. These images typically feature:
% \begin{itemize}
% \item \textbf{High-Information Cues:} Identifiable elements such as storefronts, specific architectural styles, street signs, license plates, and unique local landmarks.
% \item \textbf{Challenging Contexts:} The images are often captured from everyday perspectives (e.g., pedestrian-level, from a car window) and may include visual occlusions or non-ideal lighting, mirroring the quality of user-generated content found online.
% \end{itemize}
% Our selection from this dataset focuses on images where such fine-grained cues are pivotal for precise localization, making them ideal for testing the effectiveness of our cue-neutralizing defense.

\end{document}